\documentclass[12pt,A4]{article}
\usepackage{amsfonts}
\usepackage{mathptmx}

\begin{document}
\title{Nilpotent Classical Mechanics}
\author{Andrzej M Frydryszak \footnote{This work is partially supported by Polish KBN Grant \#
1PO3B01828} \\
Institute of Theoretical Physics, University of Wroc{\l}aw,\\ pl.
Borna 9, 50-204 Wroc{\l}aw, Poland}
\maketitle
\begin{abstract}
The formalism of nilpotent mechanics is introduced in the
Lagrangian and Hamiltonian form. Systems are described using
nilpotent, commuting coordinates $\eta$. Necessary geometrical
notions and elements of generalized differential $\eta$-calculus are
introduced. The so called $s-$geometry, in a special case when it
is orthogonally related to a traceless  symmetric form, shows some
resemblances to the symplectic geometry. As an example of an
$\eta$-system the nilpotent oscillator is introduced and its
supersymmetrization considered. It is shown that the $R$-symmetry
known for the Graded Superfield Oscillator (GSO) is present also
here for the supersymmetric $\eta$-system.
The generalized Poisson bracket for $(\eta,p)$-variables satisfies modified Leibniz rule and has nontrivial Jacobiator.
\end{abstract}
\centerline{\bf \today}
\section{Introduction}
Supersymmetric mechanics is well established theory with numerous
applications and interesting models, not all of them being just a
toy models for a supersymmetric field theory, but interesting for
themselves \cite{nico, cks, bag}. Structurally it is related to
the classical limit of theories containing fermions. Two
algebraical properties of the $\theta$-variables used in
description of the supersymmetric mechanics  are essential for
that: anticommutativity and resulting from it nilpotency. As it is
noted by Freed \cite{fred} very important issue in describing
fermions classically is nilpotency in the ring of functions of
$\theta$-variables. He notes that anticommutativity is less important.\\

In the present work we wish to study systems were only nilpotency
condition is assumed, we relax the demand to have anticommuting
(spinors) in the basic nilpotent theory. The anticommuting
variables will only appear after supersymmetrization of such a
nilpotent systems as an additional sector of the relevant model.
In the nilpotent mechanics the Pauli exclusion principle is
present only via nilpotency without other relations to spin and
statistic (anticommutativity). However having commuting and
nilpotent variables is not harmless. We have to use the
modified calculus \cite{fik}, where the usual Leibniz rule is no valid. This fact
enters in various points of the generalization. The nilpotent
mechanics in some extend indirectly shows, that for classical
description of fermions as
nilpotency as anticommutativity are equally important.\\
In the present work we firstly, in Sec.2 and Sec.3, introduce
notion of the commuting nilpotent ${\mathcal N}$-numbers and
elements of generalized differential calculus for functions of
nilpotent commuting variables. In Sec.4 we describe the $s$-geometry in the
$\mathcal{N}$-modules. In a general case $s$-forms are related to the pseudo-Euclidean metrics with arbitrary signature. Here we shall restrict to the even dimensional $s$-forms orthogonally related to the pseudo-Euclidean forms of the zero signature. This is motivated by the present formulation of the nilpotent $\eta$-mechanics. In such a case we describe a symmetry group  of the $s$-form denoted here by the $Ap(n)$.
This group is conjugated to the $O(n,n)$, however due to the
restrictions inherent in the $s$-geometry only specific matrix
realizations are possible for the elements of the $Ap(n)$ (we
shall use the $Ap(n)$ symbol to differentiate such matrices from
the generic $O(n,n)$ elements). Then, in Sec.5, we introduce the
notion of $\eta$-nilpotent mechanics in terms of the Lagrangian
and then Hamiltonian formalisms, and we show properties of the
$\eta$-Poisson bracket. This generalized bracket is linear and antisymmetric, but satisfies modified Leibniz rule and has nontrivial Jacobiator. Finally in Sec.6 we describe the
nilpotent oscillator and we give its  $N=2$ supersymmetrization.
In addition to the supersymmetry invariance there is present
interesting $R$-symmetry, already known for the Graded Superfield
Oscillator (GSO) \cite{amf18}. This symmetry intertwines two
sectors of the total system which are otherwise separately
invariant under $N=2$ SUSY.
%
\section{$\mathcal{N}$-numbers}
Our aim is to work with spaces where the coordinates are given by
nilpotent commuting $\eta$ `numbers' - an analog of the
anticommuting $\theta$ variables used in supersymmetry. Because
of the nilpotency of the $\eta$-variables we can expect some
similarities to the superanalysis, but this two formalisms are
different. The nilpotency is natural for anticommuting (odd)
elements but for commuting (even) elements it is an additional
restrictive condition. It makes for example, that the usual
Leibniz rule is not valid  and we have to modify it \cite{fik}.
We firstly introduce the notion of commuting nilpotent
$\mathcal{N}$-numbers then functions of $\mathcal{N}$-variables
and discuss , based on them, generalized differential and integral calculus.

%
Let $\bar{\mathcal{N}}_m(\mathcal{K})$ be an associative
commutative free algebra over $\mathcal{K}=\mathbb{C}$ or $\mathbb
R$ generated by the set $\{\xi^i\}_1^m$ of nilpotent elements,
$(\xi^i)^2=0$, $i=1,2,\dots,m$.
We define the unital algebra
$\mathcal{N}_m(\mathcal{K})=\mathcal{K}\oplus
\bar{\mathcal{N}}_m$. The ${\mathcal{N}}_m$ as a vector space is
isomorphic to the Grassmann algebra $\bigwedge \mathbb{V}_m $over
$m$-dimensional space $\mathbb{V}_m$, however $\mathcal{N}$ is a
commutative algebra, but still graded with the dimension equal
$2^m$. For the $\mathcal{N}_m$ we have $2^m$ monomials spanning the
algebra
\begin{equation}
1,\,
\xi^{i_1},\,\,\xi^{i_1}\xi^{i_2},\,\,\dots,\xi^{i_1}\xi^{i_2}\dots\xi^{i_k},\,\,\dots
,\,\, \xi^1\dots\xi^m
\end{equation}
Products of $\xi^{i}$ are
symmetric and indices, due to nilpotency of $\xi^{i}$, cannot
repeat therefore we will write such product of generators $\xi^i$ using strictly ordered multi-indices, namely
\begin{equation}
\xi^{I_k}=\xi^{i_1}\xi^{i_2}\dots\xi^{i_k},\quad\mbox{and}\quad \xi^{I_0}\equiv 1,\quad I_0=\emptyset,
\end{equation}
where $I_k=(i_1,\,i_2,\dots,\,i_k)$ and $i_1<i_2<\dots<i_k$. Any
element $\nu$ of $\mathcal{N}_m(\mathcal{K})$ is of the form
\begin{equation}
\nu=\sum_{k=0, I_k}\nu_{I_k}\xi^{I_k}, \quad
\nu_{I_k}\in\mathcal{K},
\end{equation}
where $\nu_{I_0}\equiv \nu_{\emptyset}$.
Analogously to the Grassmannian case, we will refer the
decomposition $\mathcal{N}=\mathcal{K}\oplus\bar{\mathcal{N}}$ to
the body and soul of an element $\nu=\nu_B+\nu_S$,
$\nu_B\in\mathcal{K}$, $\nu_S\in\bar{\mathcal{N}}$. In the finite
dimensional$\mathcal{N}_m$ any $\nu_S$ is nilpotent of some order i.e. there exists 
$n<m$ that $(\nu_S)^n=0$.
 One can also consider the infinite dimensional version of this
algebra which will be denoted by $\mathcal{N}$ (analogously to the
Banach-Grassmann algebra considered in Ref.\cite{rog, japi}). In
the following we will consider the infinite number of
algebraically independent nilpotent generators in the
$\mathcal{N}$. Let $\mathcal{M}\subset\mathcal{N}$, the
anihillator $\mathcal{M}^{\perp}$ of $\mathcal{M}$ is defined as
the following set
\begin{equation}
\mathcal{M}^{\perp}=\{\nu'\in\mathcal{N}|
\forall\nu\in\mathcal{M},\, \nu\nu'=0\}
\end{equation}
Analogously to the Grassmann algebra the $\mathcal{N}$ algebra we
shall call effective when $\bar\mathcal{N}^{\perp}=\{0 \}$.
Infinite dimensional $\mathcal{N}$ algebra is effective and in
the finite dimensional case anihillator is nontrivial and
contains the `last' element $\xi^1\dots\xi^m$. We shall use
effective $\mathcal{N}$ algebra to avoid degeneracy related to
the presence of the nontrivial anihillator. The $\mathcal{N}$
algebra will play the role of "field of numbers". Elements of
$\mathcal{N}$ with nontrivial body are invertible and in such
case inverse element is given in the following form
\begin{equation}
\nu^{-1}=\nu_B^{-1}\sum_{n=0}^{\infty}(-\nu_B^{-1}\nu_S)^n
\end{equation}
Let $\eta\in\bar\mathcal{N}$ denotes nilpotent element whose
square vanishes, $\eta^2=0$. We shall call such elements first
order nilpotents, and denote them by $\eta$ . Let
$\mathcal{D}=\{\eta\in\mathcal{N}|\eta^2=0\}\subset\bar{\mathcal{N}}$ be a set of such first order nilpotents..
Pair of the first order nilpotent elements $\eta$ and $\eta'$ we shall call
independent nilpotents if $\eta\eta'\neq 0$. Let us put, for
$k\in\mathbb{N}$
\begin{equation}
\tilde{\mathcal{D}}_k=\{(\eta_1, \eta_2,
\dots,\eta_k)\in\mathcal{D}^k=\mathcal{D}\times\mathcal{D}\times\dots\times\mathcal{D}\,|\,
\eta_1\eta_2\dots\eta_k\neq0\}
\end{equation}
Nilpotent of the first order elements do not need be monomials in $\xi^i$. For
example one can take $\eta=\nu_1\xi^1+\nu_{13}\xi^1\xi^3$, then
$\eta^2=0$. As we describe it  below,  the $\mathcal{N}_m$
algebras have matrix realizations, but since we want to use this
structure as the basics "field" of numbers,
such realizations will not be used in our approach.\\[2mm]
{\bf Examples:}
\begin{enumerate}
\item The simplest possible $\mathcal{N}$ algebra has one
nilpotent generator. Arbitrary $\nu\in\mathcal{N}_1$ has the form
\begin{equation}
\nu=\nu_0\cdot 1+\nu_1\xi, \quad \nu_0,\,\nu_1\in\mathcal{K}
\end{equation}
Matrix realization is the following
\begin{equation}
\xi=\left(
\begin{array}{ll}
0&1\\
0&0
\end{array}
\right), \quad \mbox{and} \quad \nu=\left(
\begin{array}{ll}
\nu_0&\nu_1\\
0&\nu_0
\end{array}
\right).
\end{equation}
This structure is isomorphic as an algebra to the dual numbers as
well as to the Grassmann algebra with one odd generator.
\item Matrix realization from the above example can be generalized
to the $\mathcal{N}_m$ case. The nilpotent generators
$\{\xi^i\}_1^m$ can be taken in the following form
\begin{equation}
\xi_i=\mathbb{I}_2\otimes\dots\mathbb{I}_2\otimes\xi\otimes\mathbb{I}_2
\dots\otimes\mathbb{I}_2,
\end{equation}
where $\mathbb{I}_2$ is two dimensional unit matrix and $\xi$ is
located at the $i$-th position of the tensor product. Obviously
$\xi^i\xi^j=\xi^j\xi^i\neq 0$ for $i\neq j$ and $(\xi^i)^2=0$,
$i,j=1,2,\dots,m$.
\end{enumerate}
Let us note that in the commutative algebra sum of nilpotents is
again nilpotent of some order, but for noncomutative algebras
this is not true. Obviously in the algebra $\mathcal{N}_m$ for $m>2$ we have nilpotents of order higher then one, on the other hand for our formulation nilpotents of the first order play distinguished role. In the following we will explicitly mention the degree of nilpotency when necessary, to avoid misunderstanding.
\section{$\eta$-calculus}
One can introduce the notion of differential and integral for
functions of the $\eta$-variables, in parallel to these used in
superanalysis. However, due to different properties of the
multiplication in the algebra $\mathcal{N}$ they behave
differently. As in the approach to the superalgebras and
superanalysis considered e.g in \cite{rog, japi} we shall treat
the $\eta$-variables, $\eta^2=0$, as not fixed.
\subsection{$\eta$-functions}
Let  $\eta_i$, $i=1,2,\dots, n$ be from the set of independent
nilpotent first order  elements $\eta_i\in\tilde{\mathcal{D}}_n$,
namely
\begin{equation}
(\eta^i)^2=0 \;\; \forall \,  i, \quad \eta^1\cdot \eta^2 \cdot
\dots \eta^n\neq 0
\end{equation}
and $\vec{\eta}=(\eta^1, \eta^2, \dots \eta^n)$. Moreover let, as
before, the $I_k$ denotes a strictly ordered multi-index. We shall
define the function $f(\vec{\eta})\in \mathcal{F}[\vec{\eta}]$ of
$n$ $\eta$-variables by the following expansion
\begin{equation}\label{expan}
f(\vec{\eta})=\sum_{k=0, I_k}^n f_{I_k}\eta^{I_k},
\end{equation}
where $f_{I_k}\in \mathcal{N}$ are constant elements. The expansion (\ref{expan}) gives explicitly the dependence of a function $f$ on the $\eta$-variables . When the function $f$ depends also on the
$x\in\mathcal{K}^n$, then $f_{I_k}:\mathcal{K}^n\mapsto
\mathcal{N}$, $f\in\mathcal{F}[x,\vec{\eta}]$. In the following
we will consider $f(t,\vec{\eta})=\sum f_{I_k}(t)\eta^{I_k}$,
with $t\in\mathbb{R}$. In particular one can consider the set of $\eta$-functions with coefficients in the expansion (\ref{expan}) from the $\mathcal{K}$, $f_{I_k}\in\mathcal{K}$. The such of such a functions will be denoted by $\mathcal{F}_0[x,\vec{\eta}]$. We want to stress that $\eta$-variables are treated here analogously to the $\theta$ anticommuting
variables used in superdifferential calculus.
\subsection{$\eta$-derivative}\label{eta-der}
Let us introduce the $\eta$-derivative analogously as it is done
in superanalysis, namely as a contraction defined on the basic
variables
\begin{equation}
\partial_i\eta^j=\delta_i^j, \quad \partial_i 1=0,
\end{equation}
where
\begin{equation}
\partial_j=\frac{\partial}{\partial \eta^j}
\end{equation}
By linearity over ${\mathcal{N}}$ we extend it to the
$\mathcal{F}[\vec{\eta}]$ i.e. $\partial_i (a\eta^k+b\eta^j)=a\partial_i\eta^k
+b\partial_i \eta^j$, where $a,b\in\mathcal{N}$.
Naturally,
\begin{equation}
\partial_i\partial_j=
\partial_j\partial_i
\end{equation}
Immediate conclusion from this
definition is that the conventional Leibniz rule is not valid,
instead  for $f(\vec{\eta}),\, g(\vec{\eta})\in
\mathcal{F}[\vec{\eta}]$ we
 have the following relation
\begin{equation}\label{leib}
\partial_i(f\cdot g)= \partial_i f\cdot g + f \cdot \partial_i g -
2\eta_i\partial_i f\partial_i g
\end{equation}
This is an example of generalized Leibniz rule with the so called
Leibnizian term considered in Ref. \cite{fik}. The following relations are direct
consequence of the Eq. (\ref{leib})
\begin{itemize}
\item[(i)]
\begin{equation}
\partial_i(\eta_i f)= f-\eta_i\partial_i f
\end{equation}
\item[(ii)]
\begin{equation}
[\partial_i, \eta_i]_-=1-2\eta_i\partial_i,\quad [\partial_i,
\eta_i]_+=1
\end{equation}
\item[(iii)]
\begin{equation}
\nabla_i(fg)=\nabla_i fg+f\nabla_i g,\quad \mbox{for} \quad
\nabla_i=\eta_i\partial_i\quad \mbox{(no sum)}
\end{equation}
\end{itemize}
Let us observe that one can consider "square root" of
$\eta$-functions and derivatives within the superalgebra, in the
following sense.  In the simplest case we can treat $\eta$ as a
composite object $\eta=\theta_1\theta_2$, where $\theta_i$,
$i=1,2$ are Grassmannian anticommuting variables. Taking the even
in $\theta$ functions of the form $f(\theta_1,
\theta_2)=f_0+f_{12}\theta_1\theta_2$ and second order derivative
$D=\partial^2/\partial\theta_2\partial\theta_1$ we can write
\begin{equation}
D(f(\theta_1, \theta_2)g(\theta_1, \theta_2))= Df\cdot g+f\cdot
Dg-(-1)^{|f|}(\frac{\partial}{\partial\theta_1}f\frac{\partial}{\partial\theta_2}g-
\frac{\partial}{\partial\theta_2}f\frac{\partial}{\partial\theta_1}g)
\end{equation}
where graded Leibniz rule for $\theta$-derivatives was used. For
even functions, $|f|=0$, using expansion of functions $f$, $g$
finally we get
\begin{equation}
D(f\cdot g)=Df\cdot g+f\cdot Dg-2\theta_1\theta_2 Df\cdot Dg.
\end{equation}
Expressing in above functions product of $\theta$'s by $\eta$ we
can also identify $D$ and $\partial/\partial\eta$. In this
context modified Leibniz rule for $\eta$ calculus can be
interpreted as the effect of the second order superdifferential
calculus. However, this observation does not mean that the nilpotent mechanics, we are going to introduce, can be reduced to the pseudo-mechanics or supersymmetric mechanics. Treating the $\eta$-variables as a composite - even product of the Grassmannian variables - does not automatically reduce one model to an another. 
\subsection{$\eta$-integration}
Let us use again the analogy to the superanalysis and define
$\eta$-integral by the following contractions
\begin{equation}
\int\eta_i d\eta_j=\delta_{ij}, \quad \int d\eta_i=0
\end{equation}
and by linearity over ${\mathcal{N}}$ extend it to the
$\mathcal{F}[\vec{\eta}]$. Such defined $\eta$-integral has the
following properties
\begin{itemize}
\item[(i)]
\begin{equation}
\int \vec{\eta}d\vec{\eta}=1, \quad
\vec{\eta}=\eta_1\eta_2\dots\eta_n,\quad
d\vec{\eta}=d\eta_1d\eta_2\dots d\eta_n
\end{equation}
or equivalently for strictly ordered multi-indices $I_k$
\begin{equation}
\int\eta^{I_k}d\vec{\eta}= \left\{
\begin{array}{ll}
0,&\quad k<n\\
1,&\quad k=n
\end{array}
\right.
\end{equation}
\item[(ii)]
\begin{equation}\label{intder}
\int\partial_i f(\vec{\eta})d\eta_i=0,\quad \mbox{and}\quad
\int\partial_i f(\vec{\eta})d\vec{\eta}=0,
\end{equation}
where $f(\vec{\eta})=f(\eta_1,\eta_2,\dots,\eta_n)$
\item[(iii)]the integration by part formula
\begin{equation}\label{bypart}
\left(\int fd\eta_i\right)\left(\int
g\,d\eta_i\right)=\frac{1}{2}\left( \int(\partial_i f)\cdot g
d\eta_i +  \int f\cdot(\partial_i g)d\eta_i \right)
\end{equation}
\item[(iv)] Let matrix $A$ represents permutation and scaling
transformation, $\vec{\eta}=A\vec{\eta}'$
\begin{equation}
\int f(\vec{\eta})d\vec{\eta}=(Per \,A)^{-1}\int
f(A\vec{\eta}')d\vec{\eta}',
\end{equation}
where $Per\,A$ is the permanent of the matrix $A$.
\item[(v)] $\delta$-function
\begin{equation}
\int
f(\vec{\eta})\delta(\vec{\eta}-\vec{\rho})d\vec{\eta}=f(\vec{\rho})
\end{equation}
has the following resolution
\begin{equation}
\delta(\vec{\eta}-\vec{\rho})=\prod_{i=1}^{n}(\eta_i+\rho_i)
\end{equation}
Note the plus sign in the above resolution of the $\delta$
function for $\eta$ variables (what differs it from the
$\delta$-function known from superanalysis, for anticommuting
$\theta$-variables).
\end{itemize}
The proofs of above facts are straightforward. Let us  prove the
integration by part formula. Indeed, using modified Leibniz rule
we can write
\begin{eqnarray}
\int \partial_i\left( f(\vec{\eta})\cdot
g(\vec{\eta})\right)d\eta_i&=&\int\partial_i f(\vec{\eta})\cdot
g(\vec{\eta})d\eta_i+\int f(\vec{\eta})\cdot\partial_i
g(\vec{\eta})d\eta_i\\&&- 2\int\eta_i \partial_i
f(\vec{\eta})\partial_i g(\vec{\eta})d\eta_i
\end{eqnarray}
The left-hand side of the above formula vanishes, due to the
(\ref{intder}), and
\begin{equation}
\int\eta_i\partial_i f(\vec{\eta})\partial_i g(\vec{\eta})d\eta_i=
\partial_i f(\vec{\eta})\partial_i g(\vec{\eta}),
\end{equation}
moreover $\int fd\eta_i=\partial_i f$, therefore
\begin{equation}
\int\partial_i f(\vec{\eta})\cdot g(\vec{\eta})d\eta_i=2\int
f(\vec{\eta})d\eta_i \int g(\vec{\eta})d\eta_i-\int
f(\vec{\eta})\cdot\partial_i g(\vec{\eta})d\eta_i
\end{equation}
or in in equivalent form (\ref{bypart}).
%
%
\section{The geometry of $s$-forms}\label{gcsdef}
The notion of a configuration space that we will use in the
definition of the nilpotent mechanics has two ingredients: one is
the free bimodule over commutative algebra $\mathcal{N}$, the
second is the $s$-form with values in $\mathcal{N}$ defining
geometry in such a bimodule (the geometry which is, in some
sense, compatible with nilpotency of the coordinates). Firstly
let us describe necessary notions from the $s$-geometry
\cite{amf-n1}, then we will consider $\mathcal{N}$-module which
allows to use commutative nilpotent $\eta$ variables. Main idea
behind $s$-geometry is to have a non-degenerate symmetric form,
which cannot be diagonalized, to avoid its triviality after
generalization to the coordinates which are nilpotent. Unlike the
Grassmannian case, where coordinates anticommute and natural
geometry is given by antisymmetric form, here construction is not
general-linearly covariant. We have to restrict the set of
admissible bases. Obtained description resembles the light cone
formalism for pseudo-Euclidean spaces.
Because of further applications to the nilpotent mechanics we shall
 confine ourselves to the  case when such an $s$-form can be
related to the diagonal pseudo-Euclidean one by an orthogonal
transformation. What means that we shall restrict discussion to
the $s$-forms related to pseudo-Euclidean metrics of zero
signature. We will discuss the group of symmetries of such
$s$-form in dimension $2n$, which will be denoted $Ap(n)$. It is
conjugated to the $O(n,n)$ but because of the restrictions on the
set of bases we have to use special matrix representation. It can
be obtained from the hyperbolic $c-s$ form for the
pseudo-orthogonal matrices or in the special case, from the
solution analogous to the one known for the symplectic group
$Sp(n)$.
\subsection{$s$-geometry}\label{sgeom}
We shall begin with the definition of the $s$-form in a vector
space.
 Let $\mathbb V$ be a vector space $dim\mathbb{V}=2n$, over
$\mathbb R$. The $s$-form is a $\mathbb R$-bilinear, symmetric
mapping
$$
s:\mathbb{V}\times \mathbb{V}\rightarrow \mathbb{R}
$$
which is weakly nondegenerate (i.e. if $s(v,\, v')=0$ for all $v'
\in \mathbb{V}$, then $v=0$) and strictly traceless for some set
of bases $\mathcal{B}$, where we say that metric is strictly
traceless in a basis $\{b_i\}_1^{2n}$ in $\mathbb{V}$ if
\begin{equation}
s(b_i,\, b_i)=0, \quad\forall i=1,2,\dots, 2n.
\end{equation}
Let us note here that an antisymmetric form is automatically
strictly traceless in any basis, for a symmetric form the condition
of being strictly traceless in any basis makes it trivial.
Therefore we have to restrict the set of bases admitted by
$s$-form. In the usual linear geometry terms the $s$-form would be
diagonalized in a suitable basis to the form (within the $s$-geometry  such transformation is not allowed)
\begin{equation}
diag(\underbrace{+,+,\dots,+}_p,\underbrace{-,-,\dots,-}_q)
\end{equation}
,where $p+q=2n$ and in general $p\neq q$. However, if a form $s$ can
be related to the diagonal pseudo-Euclidean by an  orthogonal
transformation, then $p=q=n$.
In the rest of this paper we shall confine to this case and
related pseudo-Euclidean metrics will be of signature zero.\\
We will say that the $s$ is given in the the standard form if
\begin{equation}\label{nat}
s(v,\, v')=\sum_{i=1}^n v_i  v_{n+i}'+v_i' v_{n+i}
\end{equation}
The name "standard" is used here in analogous sense as the
standard form of symplectic metric, moreover as it is known from
the Witt's theorem it is always possible to put nondegenerate
symmetric form in the standard form, in the so called Witt's or
hiperbolic basis \cite{koma}.  A basis $\{e_i\}_1^{2n}$ is the
$s$-admissible basis iff $s(e_i,\, e_i)=0, \forall i$. By the
$s$-space we shall understand the triple: a linear space (module)
$\mathbb{V}_{2n}$ with a $s$-form and the set of all
$s$-admissible bases
$\mathcal{B}$ i.e. $(\mathbb{V}_{2n}, s, \mathcal{B})$.\\
Two $s$-spaces $(\mathbb{V},s, \mathcal{B})$ and $(\mathbb{V'},s',
\mathcal{B}')$ are isomorphic if there exists a vector space
isomorphism $\Phi$ such that
\begin{equation}
s'(\phi(v), \phi(w))=s(v, w), \quad \mbox{where}\, v,w\in
\mathbb{V}
\end{equation}
In particular we have that $\phi(\mathcal{B})=\mathcal{B}'$. Let
$Aut_s(\mathbb{V})$ be the set of automorphisms of $(\mathbb{V},
s, \mathcal{B}$) \cite{amf-n1}.\\
Such defined $s$-form is not positive definite. In the usual
linear geometry terms it is related to the traceless hyperbolic
geometry in the light-cone formalism. We shall use "relative
length" of vectors, in the following sense: $s(b_i,b_i)$ vanishes
but we can normalize non-vanishing product of vectors from a
fixed set, e.g. $s(b_i, b_{i+n})=1$, $i=1,2,\dots, n$.
\subsection{The $Ap(n)$ group}
As we have already noted, the $s$-geometry is obtained as the restriction
of usual hyperbolic geometry in spaces with pseudo-orthogonal
meric. There are twofold restrictions here:
even dimensionality of space, smaller set of allowed bases and
transformations. The group of transformations preserving the
$s$-form (othogonally related to the pseudo-Euclidean  metric of zero signature) we shall call the $s$-plectic group and denote it $Ap(n)$. This name is
justified by some resemblances to the symplectic group $Sp(n)$
\cite{ber, lib}. To describe this symmetry group let us consider
the $s$ in the defined by (\ref{nat}) standard form. It is
represented by the following matrix
\begin{equation}\label{natural}
s=\left(
\begin{array}{ll}
0&\mathbb{I}_n\\
\mathbb{I}_n&0
\end{array}
\right) ;\quad s^2=\mathbb{I}_{2n}; \quad s^T=s; \quad
Tr(s)|_k=0, \, k=1,2, \dots,2n
\end{equation}
The $\mathbb{I}_n$ denotes $n$-dimensional unit matrix and
$Tr(s)|_k$ the trace of a principal $k\times k$ block of the
matrix $s$. Now, $D\in Ap(n)$ when
\begin{equation}\label{inv}
D^TsD=s.
\end{equation}
Because the $s$ is given by Eq.(\ref{natural}), then writing $D\in
Ap(n)$ in a block form
\begin{equation}
D=\left(
\begin{array}{cc}
P&Q\\
R&S
\end{array}
\right)
\end{equation}
with $P$, $Q$, $R$, $S$ being $n\times n$-blocks, we get that the
condition (\ref{inv}) enforces the following relations
\begin{eqnarray}\label{cond1}
R^TP&=&-P^TR, \quad R^TQ+P^TS=\mathbb{I}_n\\
S^TQ&=&-Q^TS, \quad S^TP+Q^TR=\mathbb{I}_n \label{cond2}
\end{eqnarray}
and moreover $(det D)^2=1$. Above conditions resemble the
analogous ones for the symplectic group, but unlike the symplectic
case here the sign of $det D$ depends on dimension of $s$-space).\\
Let us observe that $Ap(n)$ is isomorphic to the
pseudo-orthogonal group $O(n, n)$. If $J$ is the matrix of the
pseudo-Euclidean metric in the canonical form
\begin{equation}
J=\left(
\begin{array}{lc}
\mathbb{I}_n&0\\
0&-\mathbb{I}_n
\end{array}
\right)
\end{equation}
then $G\in O(n,n)$ if $G^TJG=J$. Because there exists the
orthogonal matrix $U$
\begin{equation}
U=\frac{1}{\sqrt{2}}\left(
\begin{array}{lr}
\mathbb{I}_n&-\mathbb{I}_n\\
\mathbb{I}_n&\mathbb{I}_n
\end{array}
\right)
\end{equation}
such that $U^{-1}sU=J$ we have isomorphism $\phi_U$ of $O(n,n)$
and $Ap(n)$ groups
\begin{equation}
D=\phi_U(G)=UGU^{-1}
\end{equation}
Let us note that matrix $s\notin O(n, n)$ and $J\notin Ap(n)$. The
matrix group $O(n, n)$ is well known and its explicit description
can be obtained from the description of the orthogonal group
$O(2n)$ by means of the exchange mapping \cite{hig}. The exchange
mapping is defined generally on block matrices by the following
formula
\begin{equation}
exc\left(
\begin{array}{ll}
A_{11}&A_{12}\\
A_{21}&A_{22}
\end{array}
\right)= \left(
\begin{array}{lr}
A_{11}^{-1}& A_{11}^{-1}A_{12}\\
A_{21}A_{11}^{-1}&A_{22}-A_{21}A_{11}^{-1}A_{12}
\end{array}
\right),
\end{equation}
where $A_{ij}$, $i,j=1,2$ are blocks of a matrix and $A_{11}$ is
assumed to be invertible. There is the "hyperbolic" version of the
$c-s$ theorem \cite{hig} which states that any pseudo-orthogonal
$G$ matrix can be written in one of the following forms (we
consider here only the case $O(n, n)$ not general $O(p,q)$)
showing explicitly four connected components of this group
\cite{fom, hig}
\begin{equation}
G=\pm\left(
\begin{array}{lr}
\tilde{c}&\pm \tilde{s}\\
\tilde{s}&\pm\tilde{c}
\end{array}
\right),
\end{equation}
where $\tilde{c}=diag(\tilde{c}_1,\dots,\tilde{c}_n)$,
$\tilde{s}=diag(\tilde{s}_1,\dots,\tilde{s}_n)$ and
$\tilde{c}^2-\tilde{s}^2=\mathbb{I}_n$. Now using the mapping
$\phi_U$ we can get general description of matrices from $Ap(n)$
\begin{equation}
\pm\left(
\begin{array}{lr}
\tilde{c}-\tilde{s}&0\\
0&\tilde{c}+\tilde{s}
\end{array}
\right),\quad\quad \pm\left(
\begin{array}{lr}
0&\tilde{c}-\tilde{s}\\
\tilde{c}+\tilde{s}&0
\end{array}
\right)
\end{equation}
$\tilde{c}-\tilde{s}=e^{-\beta}$,
$\tilde{c}+\tilde{s}=e^{\beta}$, $\beta\in\mathbb{R}$.\\
 It is
interesting to observe that for the $n=2k$ conditions
(\ref{cond1}) and (\ref{cond2}) can be solved generically in
analogous way as it is done in the case of the symplectic group
$Sp(n)$ \cite{fol}. Namely, let $n=2k$, then subsets
\begin{equation}
\mathfrak{N}=\{ \left(
\begin{array}{cc}
\mathbb{I}_n&B_1\\
0&\mathbb{I}_n
\end{array}
 \right): B_1^T=-B_1 \}
\end{equation}
\begin{equation}
\bar{\mathfrak{N}}=\{ \left(
\begin{array}{cc}
\mathbb{I}_n&0\\
B_2&\mathbb{I}_n
\end{array}
 \right): B_2^T=-B_2 \}
\end{equation}
and
\begin{equation} \mathfrak{D}=\{ \left(
\begin{array}{cc}
A&0\\
0&(A^T)^{-1}
\end{array}
 \right): A\in GL(n) \}
\end{equation}
are subgroups of $Ap(n)$ and
\begin{equation}
\bar{\mathfrak{N}}\mathfrak{D}\mathfrak{N}=\{ \left(
\begin{array}{cc}
P&Q\\
R&S
\end{array}
 \right)\in Ap(n): det P\neq 0 \}
\end{equation}
It is easy to check that $\mathfrak{N}$, $\bar{\mathfrak{N}}$ and
$\mathfrak{D}$ solve the conditions (\ref{cond1}) and
(\ref{cond2}) and form subgroups. Now taking product
\begin{equation}{\label{deco}}
\left(
\begin{array}{cc}
\mathbb{I}_n&0\\
B_2&\mathbb{I}_n
\end{array}
\right)
\left(
\begin{array}{cc}
A&0\\
0&(A^T)^{-1}
\end{array}
\right)
\left(
\begin{array}{cc}
\mathbb{I}_n&B_1\\
0&\mathbb{I}_n
\end{array}
\right) = \left(
\begin{array}{cc}
A&AB_1\\
B_2A&B_2AB_1+(A^T)^{-1}
\end{array}
\right)
\end{equation}
We see that for $A$, $det A\neq 0$ we get the following solution
\begin{equation}
P=A,\quad Q=AB_1, \quad R=B_2A, \quad S=B_2AB_1+(A^T)^{-1}
\end{equation}
again satisfying the conditions (\ref{cond1}) and (\ref{cond2}).
Therefore the $Ap(n)$ is generated by the sets $\mathfrak{D}\cup
\mathfrak{N}\cup \{s\}$ or $\mathfrak{D}\cup
\bar{\mathfrak{N}}\cup \{s\}$, where $s$ is as usual the
following matrix
\begin{equation}
s=\left(
\begin{array}{cc}
0&\mathbb{I}_n\\
\mathbb{I}_n&0
\end{array}
\right)
\end{equation}
Moreover
\begin{equation}
D^{-1}=s\,D^Ts= \left(
\begin{array}{cc}
S^T&Q^T\\
R^T&P^T
\end{array}
\right)
\end{equation}
Finally, let us note that the solution given by  the formula
(\ref{deco}) can be written by means of the exchange operator
\begin{equation}
exc\left(
\begin{array}{cc}
A^{-1}&-B_1\\
B_2&(A^T)^{-1}
\end{array}
\right) = \left(
\begin{array}{cc}
A&AB_1\\
B_2A&B_2AB_1+(A^T)^{-1}
\end{array}
\right),
\end{equation}
where $A$ is invertible and $B_i$, $i=1,2$ are antisymmetric
$n\times n$ matrices. Exchange operator maps such class of a
matrices into the $Ap(n)$ group.
%
\section{Nilpotent mechanics}
In the present section we will discuss basic issues of the
formalism of nilpotent mechanics. Lagrangian and Hamiltonian
picture is possible. We propose a generalization of the
Poisson bracket. Such an $\eta$-Poisson bracket satisfies
the modified Leibniz rule and has nontrivial Jacobiator, which  is not a subject of the Malcev identity.
\subsection{Configuration space description}
The nilpotent mechanical system will we be defined on the
configuration space being the free $\mathcal{N}$-bimodule
$\mathbb{V}_{\mathcal{N}}$ with the $\mathcal{N}$-valued $s$-form.
\begin{equation}
s:\mathbb{V}_{\mathcal{N}}\times\mathbb{V}_{\mathcal{N}}\mapsto
\mathcal{N}
\end{equation}
Since this structure is commutative, the  generalization of
definitions form the Sec. \ref{gcsdef} to the free
$\mathcal{N}$-module case, is straightforward.

We can introduce a Lagrangian with terms being analogs of the
kinetic and potential energy. This entities are
$\mathcal{N}$-valued therefore, no positive definitness can be
established. Moreover the $s$-form itself is not positive
definite either \cite{amf-n1} (even on usual $\mathbb{R}^{2n}$
space). The Lagrangian of an $\eta$-system is defined as a
$\mathcal{N}$-valued function on configuration
$\mathcal{N}$-bimodule, given by
\begin{equation}
L=\frac{m}{2}s(\dot{\eta},\dot{\eta})-V(\eta)
\end{equation}
In particular we shall consider a "quadratic" potential for the
$\eta$-oscillator.
\begin{equation}
V(\eta)=\frac{m\omega}{2}s(\eta,\eta)=\frac{m\omega}{2}s_{ij}\eta^i\eta^j.
\end{equation}
As it is known, usual variational principle can be generalized
algebraically to the case of supersymmetric systems described by
anticommuting variables. It turns out that for the nilpotent
mechanics it is possible to reach this goal too, despite the lack
of the usual Leibniz rule.\\
Let us consider the $\mathcal{N}$-valuded action of the form
\begin{equation}\label{actionint}
I[\eta^i, \dot{\eta}^i; \alpha]=\int_{t_1}^{t_2} L(\eta^i(t,
\alpha), \dot{\eta}^i(t, \alpha))dt, \quad\quad
\alpha\in\mathbb{R}
\end{equation}
One can consider generalization of the conventional \cite{car,
lan} variations
\begin{eqnarray}
\eta^i(t, \alpha)&=&\eta^i(t)+\alpha\zeta^i(t), \quad
\zeta^i(t_1)=\zeta^i(t_2)=0\\
\dot{\eta}^i(t, \alpha)&=&\dot{\eta}^i(t)+\alpha\dot{\zeta}^i(t)\\
\eta^i(t)^2&=&\zeta^i(t)^2=0, \quad
\dot{\eta}^i(t)^2=\dot{\zeta}^i (t)^2=0
\end{eqnarray}
However $\eta^i(t, \alpha)^2\neq 0$ in general. This property is
essentially different from the supersymmetric case where the first order
nilpotency is automatical. We shell consider two cases
\begin{itemize}
\item[(i)] $\eta^i$ and $\zeta^i$ are algebraically independent i.e.
$\eta^i\zeta^i\neq 0$.\\ Then $\partial^i\zeta=0$ and
$\eta^i(t,\alpha)^2=2\alpha\eta^i\zeta^i\neq 0$. Where, as before,
$\partial_i=\partial/\partial\eta^i$
\item[(ii)] $\eta_i$ and $\zeta_i$ are algebraically dependent i.e.
$\eta_i\zeta_i = 0$.\\ Hence
$\partial_i(\eta_i\zeta_i)=\zeta_i-\eta_i\partial_i\zeta_i =0$.
This means that $\zeta_i=\eta_i\partial_i\zeta_i$ and
$\zeta_i=\eta_i\eta'_i$ for some $\eta'_i\neq \eta_i$,
$\eta'_i\eta_i\neq 0$
\end{itemize}
The variation of the action is of the following form
\begin{equation}\label{varact}
\delta I=\frac{dI}{d\alpha}|_{\alpha=0}\alpha=\int_{t_1}^{t_2}
\sum_k\left(\zeta^k\frac{\partial}{\partial\eta^k}L+\frac{d}{dt}(\zeta^k)
\frac{\partial L}{\partial\dot{\eta}^k} \right)\alpha\,dt
\end{equation}
but for $\eta$-functions we have the following extension of the
time `derivative'
\begin{equation}
\frac{d}{dt}f(\vec{\eta},
\dot{\vec{\eta}},t)=\frac{\partial}{\partial t}f(\vec{\eta},
\dot{\vec{\eta}},t)+\sum_i\partial_i f(\vec{\eta},
\dot{\vec{\eta}},t)\dot{\eta}^i+\sum_i\frac{\partial}{\partial
\dot{\eta}_i} f(\vec{\eta}, \dot{\vec{\eta}},t)\ddot{\eta}^i
\end{equation}
what gives for the product of functions in Eq.(\ref{varact})the
following relation
\begin{equation}\label{var1}
\frac{d}{dt}\left(\sum_k\zeta^k \frac{\partial L}
{\partial\dot{\eta}^k}\right)=\sum_k\dot{\zeta}^k \frac{\partial
L} {\partial\dot{\eta}^k}+\sum_k\zeta^k\frac{d}{dt}(
\frac{\partial L }{\partial\dot{\eta}^k})
-2\sum_{ki}\dot{\eta}^i\eta_i\frac{\partial \zeta^k
}{\partial\eta^i}\frac{\partial }{\partial\eta^i} (\frac{\partial
L}{\partial\dot{\eta}^k})
\end{equation}
Now for algebraically independent variations last term in
(\ref{var1}) vanishes and for algebraically dependent variations
we have
\begin{equation}
\zeta^k=\eta^k\frac{\partial}{\partial\eta^k}\zeta^k
\quad\mbox{(no sum)}
\end{equation}
using this result in the variation of action (\ref{actionint}) we
finally get the analog of the Euler-Lagrange equations of motion
for both cases
\begin{eqnarray}\label{ELa}
\mbox{EL$^{(a)}$:}\quad&&\frac{\partial
L}{\partial\eta^k}-\frac{d}{dt}(\frac{\partial
L}{\partial\dot{\eta}^k})=0,\quad
\mbox{for}\quad\eta^k(\alpha)^2\neq 0\\ \label{ELb}
\mbox{EL$^{(b)}$:}\quad&&\frac{\partial
L}{\partial\eta^k}-(\frac{d}{dt}-2\dot{\eta}^k\frac{\partial}{\partial\eta^k})\frac{\partial
L}{\partial\dot{\eta}^k}=0, \quad
\mbox{for}\quad\eta^k(\alpha)^2=0
\end{eqnarray}
For Lagrangians quadratic in velocities and coordinates these two
types of equations coincide. In the present
work we will restrict ourselves to such a case.\\
It is interesting to observe that the function analogous to the
energy of the conventional system, despite the modified Leibniz
rule of the time derivative, is still "preserved" for both types
of the E-L equations of motion for the systems with quadratic
Lagrangians. Namely, we introduce the following function $E$
\begin{equation}
E(\eta, \dot{\eta})=\dot{\eta}^k\frac{\partial
L}{\partial\dot{\eta}^k}-L(\eta, \dot{\eta})
\end{equation}
and by inspection we get that
\begin{equation} \frac{d}{dt}E(\vec{\eta},
\dot{\vec{\eta}})=0.
\end{equation}
\subsection{Phase space description}
To pass to the $\mathcal{N}$-phase space description of the
system we introduce the momenta
\begin{equation}
p_k=\frac{\partial L}{\partial \dot{\eta}^k}
\end{equation}
($p_k$ are first order nilpotents for $\eta$-systems) and adapt the idea of Legendre
transformation. As in the conventional case we have to assume the
regularity of a Lagrangian, in the following sense. Namely, the
matrix $W$
\begin{equation}
W_{kl}=\frac{\partial^2 L}{\partial \dot{\eta}^k\partial
\dot{\eta}^l}
\end{equation}
should be invertible for all $(\eta^k, \dot{\eta}^k)$ (for this it
is not necessary that kinetic term is positive definite). For the
Lagrangians with the kinetic term defined within the $s$-geometry
in the $\mathcal{N}$-module we have that
$W_{ij}=\frac{m}{2}s_{ij}$, and $det(s)\neq 0$. In such a case the
EL$^{(a)}$ equations can be solved for accelerations
\begin{equation}
\ddot{\eta}_i=\sum_k {W^{-1}}^{ik}\left( \frac{\partial
L}{\partial\eta^k} - \sum_j\dot{\eta}^j\frac{\partial^2
L}{\partial \eta^j\partial \dot{\eta}^k}\right)
\end{equation}
The space $\mathcal{P}_{\mathcal{N}}$ with coordinates $(\eta^i,
p_i)$ is $4K$-dimensional, where dimension of configuration space
is even $N=2K$, due to the nonsingular $s$-form. The
generalization of the Hamiltonian function to the
$\mathcal{P}_{\mathcal{N}}$ case is the following
\begin{equation}
  H= \sum_k p_k\dot{\eta}^k - L
\end{equation}
and we get for this formalism the generalized Hamilton's equations
of motion
  \begin{eqnarray}
      \dot{p}_k = -&\frac{\partial H}{\partial \eta^k}& \\
    \dot{\eta^k} = &\frac{\partial H}{\partial p_k}&
  \end{eqnarray}
The extension of the time derivative to the phase space
$\mathcal{P}_{\mathcal{N}}$ functions $f(\eta, p)$ is given in the
following form
\begin{equation}
\frac{d}{dt}=\partial_t+\sum_k\dot{\eta}^k\partial_k+\sum_k\dot{p}_k\bar{\partial}^k,\quad
\mbox{where}\,\,\,\partial_k=\frac{\partial}{\partial\eta^k}
\,\,\,\mbox{and}\,\,\,\bar{\partial}^k=\frac{\partial}{\partial
p_k}
\end{equation}
For further convenience let us introduce the following notations
\begin{equation}
\nabla_i=\eta_i\partial_i, \quad
\bar{\nabla}^i=p^i\bar{\partial}^i, \quad \mbox{(no summation!)}
\end{equation}
It is obvious that $(\nabla_i,\bar{\nabla}^i)$ are derivations but
$(\partial_i, \bar{\partial}^i)$ are not derivations and they
satisfy the para-Leibniz rule. Using above notation this rule can
be written for the functions on $\mathcal{P}_{\mathcal{N}}$ in the
following forms (for $\eta$ derivatives, $\partial_i$)
\begin{eqnarray}
\partial_i(f\cdot g)&=&\partial_i f\cdot g +
f\cdot\partial_ig-2\eta_i\partial_if\partial_ig=\\ &=&\partial_i
f\cdot g
+f\cdot\partial_ig-\nabla_if\partial_ig-\partial_if\nabla_ig
\end{eqnarray}
and analogously for the derivatives with respect to the momenta,
$\bar{\partial}_i$.
 The  extension of the time derivative to the functions on $\eta$-phase space
 is not usual derivation, but operation which satisfies modified Leibniz
rule. We have
\begin{eqnarray}\nonumber
\frac{d}{dt}(g(\eta, p)\cdot h(\eta, p))&=&\dot{g}\cdot
h+g\cdot\dot{h}-2\sum_k\dot{\eta}^k\eta_k \partial_k
g\cdot\partial_k h
- 2\sum_k\dot{p}_k p^k \bar{\partial}^k g\cdot \bar{\partial}^k h\\
\nonumber%
&=&\dot{g}\cdot h+g\cdot\dot{h}-2\sum_k\dot{\eta}^k\nabla_k
g\cdot\partial_k h - 2\sum_k\dot{p}_k \bar{\nabla}^k g\cdot
\bar{\partial}^k h
\end{eqnarray}
 Using the equations of motion, the time derivative can be written
 as
\begin{equation}
\frac{d}{dt}f(\eta, p)=\sum_k(\bar{\partial}^k H \cdot\partial_k
f(\eta, p)-
\partial_k H\cdot\bar{\partial}^kf(\eta, p))
\end{equation}
and for the product of functions we have
\begin{equation}\label{tdm}
\frac{d}{dt}(g(\eta, p)\cdot h(\eta, p))=\dot{g}\cdot
h+g\cdot\dot{h}-2\sum_k(\bar{\partial}^k H \cdot\nabla_k
g\cdot\partial_k h -
\partial_k H \cdot\bar{\nabla}^k g\cdot \bar{\partial}^k h)
\end{equation}
Defining $\eta$-Poisson bracket as
\begin{equation}\label{poi}
\{f(\eta, p), g(\eta, p)\}_0=\sum_k(\bar{\partial}^k f(\eta, p)
\cdot\partial_k g(\eta, p)-
\partial_k f(\eta, p)\cdot\bar{\partial}^k g(\eta, p))
\end{equation}
we can finally realize the time derivative in the following form
\begin{equation}\label{tdp}
\frac{d}{dt}f(\eta, p)=\{H,f(\eta, p)\}_0
\end{equation}
This realization is consistent with the modified Leibniz rule for
the time `derivative', since using (\ref{tdp}) we have
\begin{equation}\label{pbl}
\{H,g\cdot h\}_0=\{H,g\}_0\cdot h+g\cdot\{H,h\}_0
-2\sum_k(\bar{\partial}^k H \cdot\nabla_k g\cdot\partial_k h -
\partial_k H \cdot\bar{\nabla}^k g\cdot \bar{\partial}^k h)
\end{equation}
what agrees with the formula (\ref{tdm}) and we can  write the
Hamilton's equations of motion using the bracket
$\{\cdot,\cdot\}_0$, in the form
\begin{equation}
\dot{p}_k=\{H, p_k\}_0, \quad \dot{\eta}^k=\{H, \eta^k\}_0 .
\end{equation}
\subsection{$\eta$-Poisson bracket}
Despite the similar definition (\ref{poi}) the $\eta$-Poisson
bracket for the commuting nilpotent variables differs from the
usual or graded Poisson bracket we shall call it $\eta$-Poisson
bracket or para-bracket. It is linear, antisymmetric,
satisfies modified Leibniz rule. However it does not satisfy the Jacobi
identity. We obtain nontrivial Jacobiator \cite{rowei} and moreover, the Jacobiator itself does not satisfy the Malcev identity.\\
Let $f,g,h: \mathcal{P}_{\mathcal{N}}\mapsto \mathcal{N}$ be the
$\mathcal{P}_{\mathcal{N}}$ phase space functions i.e.
$f,g,h\in\mathfrak{F}(\mathcal{P}_{\mathcal{N}})$. The
$\eta$-Poisson bracket
\begin{equation}\label{poi2}
\{f, g\}_0\equiv\sum_k(\bar{\partial}^k f \cdot\partial_k g-
\partial_k f\cdot\bar{\partial}^k g)
\end{equation}
has the following properties
\begin{eqnarray}\label{pb-pro}
&\mbox{(i)}&\quad\{f,g\}_0=-\{g,f\}_0\\
&\mbox{(ii)}&\quad\{f_1+f_2,g\}_0=\{f_1,g\}_0+\{f_2,g\}_0\\
&\mbox{(iii)}&\quad\{f,g\cdot h\}_0=\{f,g\}_0\cdot
h+g\cdot\{f,h\}_0 -2\diamondsuit(f|g,h)\\\label{lei}
&\mbox{(iv)}&\quad\sum_{cycl}\{\{f,g\}, h\}_0\}_0=J(f, g, h) \label{jaco}
\end{eqnarray}
where the para-Leibniz term is of the form
\begin{equation}
\diamondsuit(f|g,h)=\sum_k(\bar{\partial}^k f \cdot{\nabla}_k
g\cdot\partial_k h -
\partial_k f \cdot\bar{\nabla}^k g\cdot \bar{\partial}^k h).
\end{equation}
The skew-symmetric operator $J$ appearing in the Eq. (\ref{jaco})
is called Jacobiator cf. \cite{rowei} and for our $\eta$-bracket has
explicitly the following form
\begin{equation}
J(f, g, h)=2\sum_{cycl}\sum_{i}(\eta^i\{{\partial}_i f, {\partial}_i g\}\bar{\partial}^i h-p_i\{\bar{\partial}^i f, \bar{\partial}^i g\}
{\partial}_i h).
\end{equation}
To see that the Jacobi identity is violated let us consider
the following functions $f=\eta^i$, $g=p_i\eta^k$, $h=p_k p_i$, $k\neq i$.
\begin{equation}
J(\eta^i, p_i\eta^k, p_k p_i)=2 p_i
\end{equation}
The same example shows that the Malcev identity is not satisfied as well. Namely,
\begin{equation}
\{J(f, g, h), f\}\neq J(f, g, \{f, h\}\}).
\end{equation}
%
\section{Supersymmetric  nilpotent oscillator}\label{sno}
In the present section we shall consider the simple system of the
two dimensional nilpotent $\eta$-oscillator (i.e. two is the
lowest dimension of the nontrivial $s$-space). Then its
supersymmetrization within the $N=2$ SUSY multiplet. The
nilpotent oscillator has its very close counterpart in the
fermionic oscillator which is described using anticommuting
variables and antisymmetric form $\epsilon^{ij}$. Here again the
simplest example is two dimensional. Using the $N=2$ SUSY
multiplet we supersymmetrize the the fermionic oscillator and
then combine both independently supersymmetric systems into one.
Such a new total system exhibits the $R$-symmetry known from the
GSO \cite{amf18}. We also comment the issue of the `parity
duality' between systems.
\subsection{Nilpotent harmonic $\eta$-oscillator}
As the simplest system realizing the nilpotent mechanics in
$\mathcal{N}$-module $\mathbb{V}_{\mathcal{N}}$ with nontrivial
$s$-geometry, let us consider the $\eta$-oscillator in two
dimensions. It is defined by the Lagrangian of the form
\begin{equation}
\mathcal{L}=\frac{m}{2}s_{ij}\dot{\eta}^i\dot{\eta}^j-
\frac{m\omega}{2}s_{ij}\eta^i\eta^j, \quad i,j=1,2
\end{equation}
The properties of the $s$-form provide that this Lagrangian is
nontrivial and gives equations of motion analogous to the
conventional case.
\begin{equation}\label{eqnho}
\ddot{\eta}_i=-\omega^2\eta_i
\end{equation}
The passage to the phase space description yields the
$\eta$-Hamiltonian containing $s$-form on the
$\mathcal{P}_{\mathcal{N}}$
\begin{equation}
H=\frac{1}{2m}(s^{-1})^{ij}p_ip_j+\frac{m\omega^2}{2}s_{ij}x^ix^j
\end{equation}
and generalized Hamilton's equations of motion
\begin{equation} \left\{
\begin{array}{rcl}
\dot{x}^i&=&\frac{1}{m}(s^{-1})^{ij}p_j \\
\dot{p}_i&=&-m\omega^2 s_{ij}x^j
\end{array}
\right.
\end{equation}
\subsection{Supersymmetric nilpotent oscillator}
To supersymmetrize the nilpotent oscillator we introduce
analogously to the usual bosonic oscillator case, the multiplet of
functions with values in the $\mathcal{N}$ algebra for commuting
components and with values in the odd part of the Banach-Grassmann
algebra $\mathcal{Q}_1$ \cite{japi} for anticommuting components.
Therefore we have
\begin{equation}\label{multip}
  \eta_i \mapsto (\eta_i(t), \psi^{\alpha}_i(t), F_i(t)),
\end{equation}
where the parity of functions is: $|\eta_i|=|F_i|=0$,
$|\psi^{\alpha}_i|=1$. Now the supersymmetric Lagrangian has
fully analogous form to the usual supersymmetric model, namely
\begin{equation}\label{laeta}
\mathcal{L}_{\eta}=\frac{1}{2}s^{ij}(\dot{\eta}_i \dot{\eta}_j+F_i
F_j)+\frac{1}{2}\delta_{\alpha\beta}s^{ij}\psi^{\alpha}_i\dot{\psi}^{\beta}_j-
\frac{\omega}{2}s^{ij}(\eta_iF_j+F_i\eta_j-\epsilon_{\alpha\beta}\psi^{\alpha}_i
\psi^{\beta}_j)
\end{equation}
Action defined by this Lagrangian is invariant under the
following supersymmetry transformations (obviously we have usual
$s$-geometry $Ap(2)$ invariance as well)
\begin{eqnarray}\label{susyeta}
\delta_{\varepsilon}\eta_i=\varepsilon_{\alpha}\psi^{\alpha}_i\\
\delta_{\varepsilon}\psi^{\alpha}_i=\epsilon^{\alpha\beta}\varepsilon_{\alpha}F_i
-\varepsilon_{\alpha}\dot{\eta}_i\\
\delta_{\varepsilon}F_i=
\varepsilon^{\alpha}\epsilon_{\alpha\beta}\dot{\psi}^{\beta}_i
\end{eqnarray}
The Lagrangian $\mathcal{L}_{\eta}$ transforms by the total time
derivative term, like in the usual supersymmetric case. Let us
note that $(\eta_i)^2=0$ but $(\delta_{\varepsilon}\eta_i)^2\neq
0$. The equations of motion that one obtains for the present
system have the form
\begin{eqnarray}\label{eqmeta}
\omega\eta_i-F_i=0\\
\omega\psi^{\alpha}_i-\epsilon^{\alpha\beta}\dot{\psi}_{\beta
i}=0\\
\omega F_i+\ddot{\eta}_i=0,
\end{eqnarray}
and after eliminating the auxiliary function $F_i$ we get that
$\ddot{\eta}_i=-\omega^2\eta_i$.

\subsection{Graded nilpotent oscillator and $R$-symmetry}
As it is known \cite{amf18}, for the even dimensional
configuration space it is possible to form a composite system of
supersymmetric bosonic oscillator and  supersymmetric fermionic
one. This system is separately supersymmetric in both parts, but
it exhibits also a nontrivial so called $R$-symmetry \cite{amf18}
which intertwines both parts of such a composite Lagrangian.
Moreover it relates kinetic terms with potential ones. We will
see here that this symmetry survives also in the generalization
to the graded nilpotent oscillator.\\
To fix the notation, let us firstly recall the Lagrangian of the
supersymmetric fermionic oscillator \cite{amf18}. Here we depart
from the anticommuting coordinates of the basic configuration
space $\psi_i$ and associate the following multiplet of functions
\begin{equation}\label{multipf}
\psi_i \mapsto (\psi_i(t), \eta^{\alpha}_i(t), A_i(t)),
\end{equation}
where the parity of functions is: $|\psi_i|=|A_i|=1$,
$|\eta^{\alpha}_i|=0$. Now the $N=2$ SUSY transformations on this
multiplet have the following form \cite{amf18}
\begin{eqnarray}\label{susypsi}
\delta_{\varepsilon}\psi_i=\varepsilon_{\alpha}\eta^{\alpha}_i\\
\delta_{\varepsilon}\eta^{\alpha}_i=\epsilon^{\alpha\beta}\varepsilon_{\alpha}A_i
-\varepsilon_{\alpha}\dot{\psi}_i\\
\delta_{\varepsilon}A_i=
\varepsilon^{\alpha}\epsilon_{\alpha\beta}\dot{\eta}^{\beta}_i
\end{eqnarray}
As before we can write the Lagrangian yielding the action
invariant under SUSY transformations (\ref{susypsi}), namely
\begin{equation}\label{lapsi}
\mathcal{L}_{\psi}=-\frac{1}{2}\epsilon^{ij}(\dot{\psi}_i
\dot{\psi}_j+A_i
A_j)+\frac{1}{2}\delta_{\alpha\beta}\epsilon^{ij}\eta^{\alpha}_i\dot{\eta}^{\beta}_j-
\frac{\omega}{2}\epsilon^{ij}(\psi_iA_j+A_i\psi_j-\epsilon_{\alpha\beta}\eta^{\alpha}_i
\eta^{\beta}_j)
\end{equation}
The equations of motion for the supersymmetric fermionic
oscillator are of the following form
\begin{eqnarray}\label{eqmpsi}
\omega\psi_i+A_i=0\\
\omega\eta^{\alpha}_i+\epsilon^{\alpha\beta}\dot{\eta}_{\beta
i}=0\\
\omega A_i-\ddot{\psi}_i=0,
\end{eqnarray}
and after eliminating the auxiliary function $F_i$ we get that
$\ddot{\psi}_i=-\omega^2\psi_i$.\\
Let us observe that there is a kind of an informal `parity
duality' between the $N=2$ supersymmetric nilpotent oscillator
and $N=2$ supersymmetric fermionic oscillator, in the sense that
\begin{eqnarray}
(\eta_i)^2=0, \quad &\eta_i \longleftrightarrow \psi_i&, \quad
(\psi_i)^2=0\\
&\psi^{\alpha}_i \longleftrightarrow \eta^{\alpha}_i&\\
&F_i \longleftrightarrow A_i&\\
&s^{ij} \longleftrightarrow \epsilon^{ij}&
\end{eqnarray}
But such relation is not working on the level of derivatives with
respect to the nilpotent co-ordinates. We cannot directly relate
\begin{equation}
\frac{\partial}{\partial\psi_i}
\end{equation}
which is $\mathbb{Z}_2$ graded derivative and
\begin{equation}
\frac{\partial}{\partial\eta_i}
\end{equation}
which is para-derivative, with nontrivial para-Leibniz term.\\
Now we can combine two independently invariant under $N=2$ SUSY
Lagrangians (\ref{laeta}) and (\ref{lapsi}) and form the graded
nilpotent oscillator
\begin{equation}\label{gnl}
\mathcal{L}=\mathcal{L}_{\eta}+\mathcal{L}_{\psi}
\end{equation}
It turns out that such a system is invariant under the
generalization of the $R$-symmetry introduced in
Ref.\cite{amf18}. This transformation is parametrized by the odd
parameter $\xi$. For the $\eta$-multiplet (\ref{multip}) it takes
the form
\begin{eqnarray}\label{reta}
\delta_{\xi}\eta_i&=& \xi s_{ij}\frac{1}{\omega}\dot{\psi}^j\\
\delta_{\xi}\psi^{\alpha}_{i}&=&-\xi s_{ij}(\epsilon^{\alpha\beta}
\eta_{\beta}^j-\frac{1}{\omega}\delta^{\alpha\beta}
\dot{\eta}_{\beta}^j)\\
\delta_{\xi}F_i&=&\xi
s_{ij}(\frac{1}{\omega}\dot{A}_j+2\dot{\psi}_j)
\end{eqnarray}
and for the $\psi$-multiplet (\ref{multipf})
\begin{eqnarray}\label{rpsi}
\delta_{\xi}\psi_i&=& -\xi \epsilon_{ij}\frac{1}{\omega}\dot{\eta}^j\\
\delta_{\xi}\eta^{\alpha}_{i}&=&-\xi
\epsilon_{ij}(\epsilon^{\alpha\beta}
\psi_{\beta}^j+\frac{1}{\omega}\delta^{\alpha\beta}
\dot{\psi}_{\beta}^j)\\
\delta_{\xi}A_i&=&-\xi
\epsilon_{ij}(\frac{1}{\omega}\dot{F}_j-2\dot{\eta}_j)
\end{eqnarray}
The full Lagrangian (\ref{gnl}) transforms by total time
derivative
\begin{eqnarray}\label{rinv}
\mathcal{L}=&-&2\omega \frac{d}{dt}\left( \psi_i\eta^i \right)+
\frac{d}{dt}\left( F_i\psi^i-\eta^i
A_i-\eta^{i\alpha}\epsilon_{\alpha\beta}\psi^{\beta}_i
\right)\\\nonumber%
&+&\frac{1}{\omega}\frac{d}{dt}\left(
\dot{\eta}_i\dot{\psi}^i +F_i
A^i+\eta^{\alpha}_i\dot{\psi}^{\alpha i}
-\dot{\eta}^{\alpha}_i\psi^i_{\alpha} \right)
\end{eqnarray}
It is worth noting, that in each term of above total time
derivative, there are contributions or cancellations coming from
both parts of the Lagrangian (\ref{gnl}). The $R$-symmetry mixes
both sectors defined by multiplets $(\eta_i(t),
\psi^{\alpha}_i(t), F_i(t))$ and $(\psi_i(t), \eta^{\alpha}_i(t),
A_i(t))$. Moreover, it also mixes the kinetic and potential
terms. The $R$-symmetry fixes the scale between both parts of
these otherwise, independent systems.

\section*{Conclusions}
In the present work we have discussed the possibility of
constructing  formalism based on the nilpotent commuting
coordinates. Such generalization is tempting, having in mind the
supersymmetric systems, where the classical formalism is based on
the superspaces and supergeometry with anticommuting nilpotent
variables. There $\mathbb{Z}_2$-graded structures provide the
generalized description and
allow to study bosons and fermions in unified language.\\
The present approach shows that we can consider nontrivial
systems defined by nilpotent commuting coordinates, but their
description is less natural then that of graded systems given by
anticommuting coordinates. In his book \cite{fred} Freed
considers, in the context of supermanifolds, the idea of "nilpotent
cloud" or "nilpotent fuzz" around points stressing nilpotency as
its primary property. Here we tried to find some consequences of
using solely nilpotency without anticommutativity. Let us observe
that in the simplest possible case of the $\mathcal{N}_1$ algebra,
from the point of view of the set of functions
$f:\mathcal{N}_1\mapsto \mathcal{N}_1$ there is no difference
between $\mathbb{Z}_2$-graded version and commutative version of
the algebra. One can say that in both cases there is no difference
between these two kinds of the "nilpotent fuzz". But when introducing the
differential calculus in the $\mathbb{Z}_2$-graded case we obtain
superderivative anticommuting with supervariables, in the
commutative nilpotent case we have differential calculus with the
deformed
Leibniz rule and both constructions are different.\\
We have also shown that $\eta$-nilpotent systems can be
supersymmetrized and exhibit interesting (super)symmmetries.
Phase space description of $\eta$-mechanical systems is not ordinary.
The analysis given in the present work shows that it is not obvious
how to define generalization of the Poisson or even Poisson-Malcev
\cite{vers}structure. \\
As we have mentioned in the Sec.\ref{eta-der} one can consider
$\eta$ variables as a composite objects $\eta=\theta_i\theta_j$,
with $\theta$'s being Grassmannian variables, but from the point of
view of the nilpotent mechanics reduction of the $\eta$-model to
the pseudo-mechanical (model with Grassmannian co-ordinates) or supersymmetric one is not automatical, and in general, not useful. For example the $\eta$-Lagrangian written in terms of the $\theta$-variables does not define pseudo-mechanical system of any interest.\\

\section*{Acknowledgements}
The author wants to thank Andrzej Borowiec for discussions and
helpful comments.
%

\end{document}